\documentclass[10pt,conference]{IEEEtran}

\usepackage[utf8]{inputenc}
\usepackage{soul}
\usepackage{xcolor}
\usepackage{xspace}
\usepackage{url}
\usepackage{graphicx}
\usepackage{listings,multicol}
\usepackage[caption=false]{subfig}  
\usepackage[export]{adjustbox}
\usepackage{amsmath}
\usepackage{amssymb}
\usepackage{textcomp}
\usepackage{breakurl} 
\usepackage{listings}
\usepackage{hyperref}
\usepackage{cleveref}
\usepackage{algorithm}
\usepackage[noend]{algpseudocode}
\usepackage{booktabs}
\usepackage{multirow}
\usepackage{enumerate}
\usepackage{array}
\usepackage{balance}

\newcommand{\ie}{\emph{i.e.,}\xspace}
\newcommand{\eg}{\emph{e.g.,}\xspace}

\newcommand{\libraries}{\textit{RLE}\xspace}
\newcommand{\direct}{\textit{DRE}\xspace}
\newcommand{\jaccard}{\textit{Jaccard}\xspace}
\newcommand{\baseline}{\textit{Baseline}\xspace}

\newcommand{\numSurvey}{eight\xspace}

\newcommand{\secpart}[1]{\subsection{#1}}

\makeatletter
\newcommand\footnoteref[1]{\protected@xdef\@thefnmark{\ref{#1}}\@footnotemark}
\makeatother

\makeatletter
\newcommand{\linebreakand}{%
  \end{@IEEEauthorhalign}
  \hfill\mbox{}\par
  \mbox{}\hfill\begin{@IEEEauthorhalign}
}
\makeatother

\title{So Much in So Little: Creating Lightweight Embeddings of Python Libraries}

\author{
\IEEEauthorblockN{Yaroslav Golubev}
\IEEEauthorblockA{\textit{JetBrains Research}\\
Belgrade, Serbia \\
yaroslav.golubev@jetbrains.com}

\and

\IEEEauthorblockN{Egor Bogomolov}
\IEEEauthorblockA{\textit{JetBrains Research}\\
Paphos, Cyprus \\
egor.bogomolov@jetbrains.com}

\linebreakand

\IEEEauthorblockN{Egor Bulychev*}
\IEEEauthorblockA{\textit{Huawei} \\
Moscow, Russia \\
egor.bulychev@huawei.com}

\and

\IEEEauthorblockN{Timofey Bryksin}
\IEEEauthorblockA{\textit{JetBrains Research}\\
Limassol, Cyprus \\
timofey.bryksin@jetbrains.com}
}

\begin{document}

\maketitle

\begin{abstract}

In software engineering, different approaches and machine learning models leverage different types of data: source code, textual information, historical data. An important part of any project is its dependencies. The list of dependencies is relatively small but carries a lot of semantics with it, which can be used to compare projects or make judgements about them. 

In this paper, we focus on Python projects and their PyPi dependencies in the form of \textit{requirements.txt} files. We compile a dataset of 7,132 Python projects and their dependencies, as well as use Git to pull their versions from previous years. Using this data, we build 32-dimensional embeddings of libraries by applying Singular Value Decomposition to the co-occurrence matrix of projects and libraries. We then cluster the embeddings and study their semantic relations. 

To showcase the usefulness of such lightweight library embeddings, we introduce a prototype tool for suggesting relevant libraries to a given project. The tool computes project embeddings and uses dependencies of projects with similar embeddings to form suggestions. To compare different library recommenders, we have created a benchmark based on the evolution of dependency sets in open-source projects. Approaches based on the created embeddings significantly outperform the baseline of showing the most popular libraries in a given year. We have also conducted a user study that showed that the suggestions differ in quality for different project domains and that even relevant suggestions might be not particularly useful. Finally, to facilitate potentially more useful recommendations, we extended the recommender system with an option to suggest rarer libraries.

\end{abstract}

\section{Introduction}

\renewcommand*{\thefootnote}{\fnsymbol{footnote}}
\footnotetext[1]{The work was carried out when the author worked at JetBrains.}
\renewcommand*{\thefootnote}{\arabic{footnote}}

A lot of tasks in software engineering research rely on building representations of different entities: code snippets, files, or entire projects. Such representations, or \textit{embeddings}, can be used in searching for similar repositories~\cite{sosed2020}, suggesting method names~\cite{alon2018codeseq}, or code completion~\cite{svyatkovskiy2018pythia}. However, software engineering contains not only the code itself: it also leverages textual information (\eg READMEs, issues, docstrings), historical data (\eg project's age and evolution), collaboration data, licensing information, and much more. For instance, the collaboration data can be used to assign reviewers~\cite{thongtanunam2015review} or track learning in software teams~\cite{kovalenko2020codestyle}.

One more source of valuable information about the project is its dependencies. In a way, dependencies can be viewed as a skeleton of a project. Oftentimes, one glance at them can give you a pretty good idea of the project's domain: web, machine learning, media, etc. Thus, dependencies can potentially be used to compare or differentiate between projects.

In this paper, we use dependencies to create embeddings of libraries and projects, and showcase their possible usefulness by suggesting potentially interesting libraries to developers of a given project. We compiled a dataset of \texttt{requirements.txt} files of 7,132 Python projects, as well as pulled up their versions from previous years where it is possible. We have applied Singular Value Decomposition (SVD)~\cite{SVD} to the co-occurrence matrix of projects and their dependencies to obtain 32-dimensional vectors of libraries. In contrast to the approaches based on the study of in-code imports~\cite{import2vec}, our technique does not require any code processing and can be considered lightweight.

Firstly, we studied the meaningfulness of the obtained embeddings by clustering them and manually inspecting the resulting clusters. We found that the clusters themselves and their relative positions reflect semantic differences between libraries, with web- and application-related libraries laying further away from machine learning and data science fields.

We also developed a system that employs the obtained embeddings to suggest possible relevant libraries to a given project. To search for similar projects, we consider several ways to represent projects based on their dependencies. To evaluate the performance of different approaches, we also created a benchmark from the collected \textit{versions} of requirements files. The benchmark measures the models' accuracy in suggesting libraries to the projects in a given year that were actually added in the next year. The best results using the extracted embeddings demonstrate the MRR (Mean Reciprocal Rank)~\cite{mrr} of 0.189, while the baseline of suggesting the most popular libraries only shows the MRR of 0.144.

To further evaluate the usefulness of the approach, we also conducted a user study. We selected a repository from each of the five domains of Python projects~\cite{lin2016empirical}: NLP, Web, Media, Data Processing, and Scientific Calculations, and collected the suggestions of different models. We then showed the suggestions to \numSurvey Python developers and asked them to evaluate their usefulness and relevance. The results suggest that for certain domains, recommendations are better than for others, and that oftentimes the suggestions are relevant but not useful, in the sense that they might not be easily implemented in a given project. This can happen, when, for example, the approach suggests an analogous popular framework.

Finally, we propose another possible way of suggesting relevant libraries called the \textit{exploration}, which consists in facilitating the recommendation of rare libraries and therefore possibly helping developers discover new tools.

The main contributions of this paper are:

\begin{itemize}
    \item \textbf{Approach}. We proposed a novel approach to create lightweight embeddings of Python libraries and projects based on their co-occurence, and used these embeddings to create a system for recommending relevant libraries for Python projects.
    \item \textbf{Dataset and benchmark}. We collected a dataset of 7,132 Python projects and their dependencies, and used them to compile a benchmark for the libraries recommendation task. The benchmark contains 4,678 entries that consist of a project's current libraries and the set of libraries added within the next year.
    \item \textbf{Evaluation}. We compared several different approaches on the benchmark, with the embeddings-based models demonstrating the best performance. Also, we conducted a user evaluation that showed that the quality of predictions varies by domains and that the suggestions might be relevant and yet be difficult to integrate in the project that they are suggested for. 
\end{itemize}

The collected dataset, the benchmark based on it, the source code of our approach, and the CLI tool for recommending relevant libraries are available online: \url{https://github.com/JetBrains-Research/similar-python-dependencies}.

The rest of the paper is organized as follows. \Cref{sec:rw} describes the recent works related to software libraries and entity embeddings. In \Cref{sec:dataset}, we describe the collection of our dataset and discuss its nature. \Cref{sec:approach} describes in detail the creation of the embeddings and the studying of their semantics using clustering, while \Cref{sec:recom} describes the proposed system for recommending relevant libraries, the comparison of different models on the compiled benchmark, as well as the user study and their results. In \Cref{sec:threats}, we discuss possible threats to the validity of our study, and, finally, in \Cref{sec:conclusion}, we draw our conclusions and discuss possible directions of the future work. 
\section{Related Work}
\label{sec:rw}

Over the last decade, machine learning methods, and neural networks in particular, have been applied in various fields, such as natural language processing~\cite{belinkov-glass-2019-analysis}, image processing~\cite{krizhevsky2012}, machine translation~\cite{Bahdanau2015NeuralMT}, recommendation systems~\cite{youtube2016}. Software engineering (SE) domain is not an exception: different neural network models show state-of-the-art results in many SE tasks, \eg method name prediction~\cite{fernandes2019structured}, bug localization~\cite{Hellendoorn2020GlobalRM}, commit message generation~\cite{liu2018nmt}.

Depending on the task, the input of ML models can vary greatly. However, any input should be transformed into a numeric form in order to apply neural networks. A projection of the input into numeric vectors, commonly called \emph{embedding}, can be trained alongside with training the network for a specific task. Alternatively, approaches like word2vec~\cite{word2vec} build generic embeddings that can be further utilized in various downstream tasks. The latter is of interest for us, since general approaches are more suitable for our goal of exploring the information contained in project dependencies.

Two common ways to build general embeddings are modifications of word2vec~\cite{word2vec} (\eg FastText~\cite{fasttext}, GloVe~\cite{glove}) and language models (LMs)~\cite{elmo}. In order to work, both approaches need sequences of items (\eg words in text, queries in search sessions). When the sequential information is missing or is not reliable, as in the case of project dependencies, item embeddings can be built based on the co-occurrence matrix. A popular technique to transform a matrix of word-text co-occurrence is Singular Value Decomposition (SVD)~\cite{SVD}.

Theeten et al.~\cite{import2vec} studied the idea of building embeddings for software libraries. The authors collected a dataset of projects in Java, JavaScript, and Python, extracted dependencies from source code files, and trained word2vec-like model on the sets of imports contained in the same file. This approach relies on parsing the dependencies from code, and then training a small neural network to obtain embeddings. Theeten et al. also introduce a task of contextual code search: given a set of anchor libraries, they suggest similar libraries based on their embeddings. In order to evaluate the model's performance for this task, they use a somewhat artificial setup: for a set of libraries, or context, the authors identify projects that contain it and try to predict the rest of the dependencies of these projects.

Aside from building library embeddings, researchers explored the library migration graphs~\cite{teyton2012mining}, studied the way developers handle vulnerable dependencies~\cite{Pashchenko2018vulnerable} and the evolution of dependencies~\cite{bavota2013evolution}. We believe that all of these tasks can benefit from approaches that automate the extraction of semantic information for software libraries.

\section{Dataset}
\label{sec:dataset}

Our approach to retrieving semantic information for software libraries relies on projects and the lists of their dependencies. In our work, we chose Python as a target language. The first reason for this is Python's popularity and the second reason is that it has very convenient package management systems. To keep the approach simple, we focused on the \textit{pip} package installer~\cite{pip} and the \textit{PyPi} index~\cite{pypi}.

\textit{pip} is a popular package installer for Python that supports so-called \textit{requirement files}. These files must adhere to certain simple formatting rules and contain dependencies that are necessary to run the given project. If such file is present, the user can then run a single command to automatically install or update all the necessary packages via \textit{pip}.

\subsection{Data collection}

Since our method relies only on the lists of dependencies, it does not require a lot of storage space or computational resources. Thus, we decided to compile our own dataset with fresh data. As a starting point, we took GHTorrent~\cite{Gousi13}, a large collection of GitHub data, more specifically, their dump of July of 2020~\cite{ghtorrent_dumps}. To process it, we used a tool called PGA-create~\cite{pga_create} that was previously used to create a Public Git Archive~\cite{markovtsev2018public}. This tool processes the SQL dump to create a CSV with a list of projects that allows for their convenient filtering. We selected projects with at least 50 stars, which allows us to only work with at least moderately popular projects.

Then, we ran each of the obtained projects through GitHub API and filtered out the following repositories:

\begin{itemize}
    \item repositories that now redirect to other repositories in the dataset (since some repositories were moved or merged after the GHTorrent dump was carried out);
    \item repositories, the main language of which is not Python (to discard possible cases where Python is used as a supporting tool only);
    \item forks.
\end{itemize}

This process resulted in 26,072 projects that we consider to be of interest to us. \textit{pip} requirements files are traditionally named \textit{requirements.txt} and saved in the root directory of the project. Stemming from this convention, we detected all the repositories that have such a file in the root of the HEAD branch, and cloned them. This resulted in 7,876 repositories cloned in November of 2020.

In order to analyze the evolution of project libraries, we traversed the history of each project and collected versions of requirements files from Novembers of previous years, dating back to 2011. Naturally, the vast majority of projects are younger than that, so older temporal slices have fewer projects and fewer requirements files.
The very final step of the data collection is to parse the collected requirement files. We utilized a Python library called Requirements Parser~\cite{requirements_parser}, which parses the file automatically and allows for a convenient iteration over the dependencies. We skipped empty files and files with all the requirements commented out. The final tally of the projects with parsed dependencies used in this study is presented in \Cref{table:dataset}.

\begin{table}[h]
\centering
  \caption{The amount of projects with their dependencies \\ in different temporal slices in the dataset.}
  \label{table:dataset}
  \begin{tabular}{ccc}
    \toprule
    \textbf{Year} & \textbf{Projects} & \textbf{Unique libraries}\\
    \midrule 
    \textbf{2011} & 103 & 272 \\
    \textbf{2012} & 363 & 985\\
    \textbf{2013} & 938 & 1,196\\
    \textbf{2014} & 1,616 & 1,896 \\
    \textbf{2015} & 2,525 & 2,699\\
    \textbf{2016} & 3,392 & 3,573\\
    \textbf{2017} & 4,323 & 4,367\\
    \textbf{2018} & 5,248 & 5,323\\
    \textbf{2019} & 6,296 & 6,285 \\
    \textbf{2020} & 7,132 & 7,168\\
    \bottomrule
  \end{tabular}
\end{table}

\subsection{Libraries distribution}
\label{sec:distribution}

\begin{figure}[h]
  \centering
  \includegraphics[width=\columnwidth]{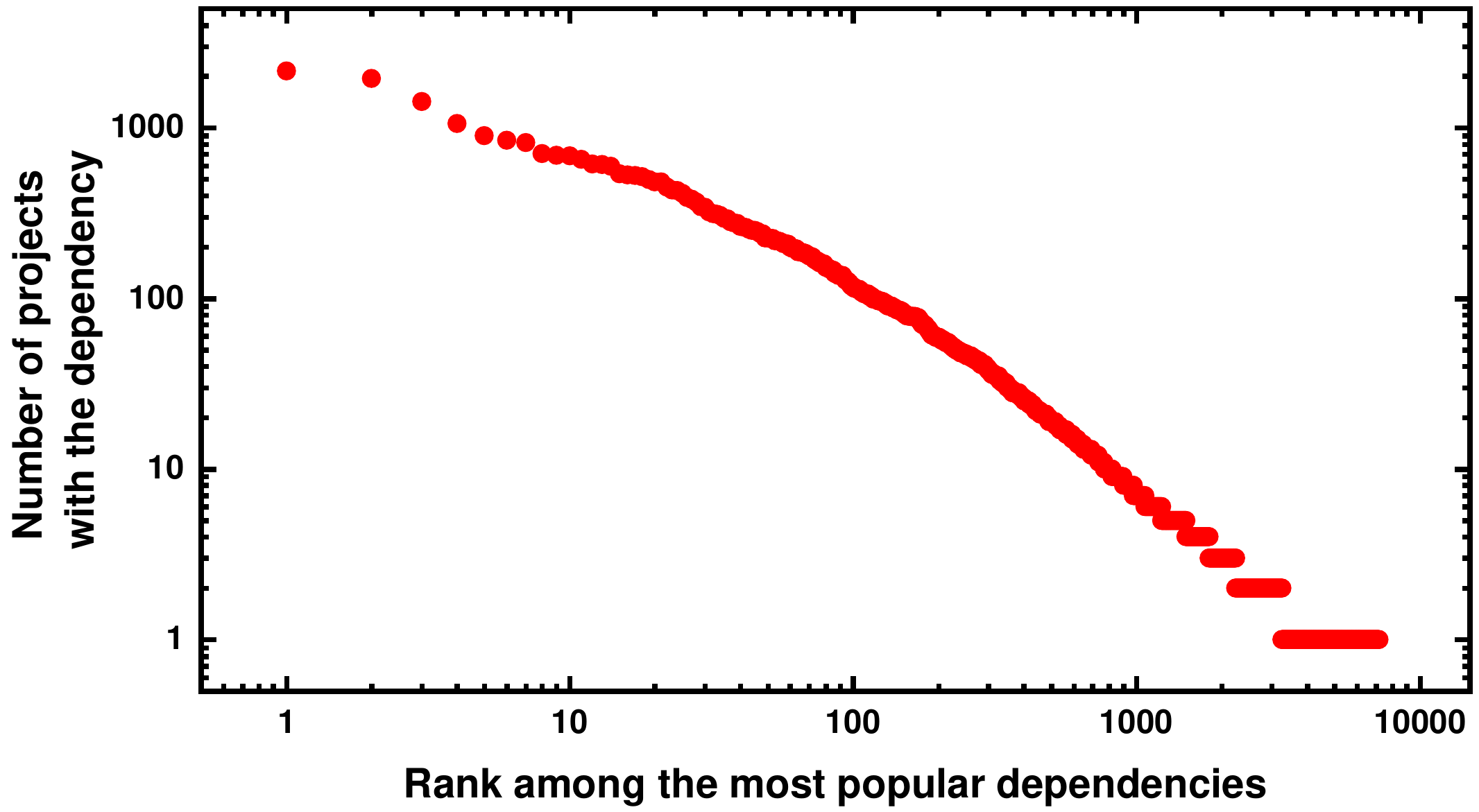}
  \caption{The distribution of the dependencies among the projects in 2020. \\ The scale is logarithmic.}
  \label{fig:dependencies}
\end{figure}

The collected dataset shows an extremely imbalanced distribution of libraries among the projects. Such a distribution among the 7,132 projects in 2020 is shown in \Cref{fig:dependencies}. Following the work by Theeten et al.~\cite{import2vec}, we conclude that the distribution of libraries roughly follows the Zipf's law. 89.3\% of all the dependencies are only encountered in 10 projects or less, and as much as 54.6\% appear only once. On the other end of the spectrum, there is a handful of very popular libraries. Top 10 most popular libraries are listed in \Cref{table:top10}.\footnote{A lot of PyPi libraries will be mentioned in this paper. In order not to create dozens of extra footnotes or references, let us say beforehand that you can find the information about them at https://pypi.org/project/name\_of\_the\_library/}

\begin{table}[t]
\centering
  \caption{Ten most popular libraries in the projects from 2020.}
  \label{table:top10}
  \begin{tabular}{c c c}
    \toprule
    \textbf{Library} & \textbf{\# of projects} & \textbf{\% of all projects}\\
    \midrule 
    requests & 2,158 & 30.3\% \\
    numpy & 1,952 & 27.4\% \\
    six & 1,428 & 20\% \\
    scipy & 1,063 & 14.9\% \\
    pyyaml & 902 & 12.6\% \\
    matplotlib & 845 & 11.8\% \\
    pillow & 818 & 11.5\% \\
    pandas & 705 & 9.9\% \\
    python-dateutil & 690 & 9.7\% \\
    tqdm & 687 & 9.6\% \\
    \bottomrule
  \end{tabular}
\end{table}

\section{Embeddings of libraries}
\label{sec:approach}

To use project dependencies as a source of information for SE tasks, we need to represent them in a numerical way. The dependency data is very sparse: each project depends only on a handful of all libraries. Also, the distribution of dependency frequencies is very skewed, with a small amount of very popular libraries and a long tail of libraries which occur as a dependency in few projects (see \Cref{fig:dependencies}).

Similar issues arise in the NLP (Natural Language Processing) domain, where each document contains only a small amount of all words, and frequently used words constitute a small part of vocabulary. A common way to resolve both issues is to use \emph{embeddings}, dense numerical vectors that contain information about word semantics. Following this idea, we build embeddings of libraries and further analyze that they are indeed meaningful and useful in the SE domain.

\secpart{Building library embeddings}
\label{sec:svd}

Following the early work in the word embeddings domain~\cite{steinberger2005svd}, we build the vectors with SVD (Singular Value Decomposition). As an alternative, we considered other popular approaches to create embeddings such as language models~\cite{elmo} or adaptation of \emph{word2vec}~\cite{word2vec}. 
However, both approaches require sequential information to operate, whereas items in requirement files do not have any particular order.

To build embeddings with SVD, we construct a co-occurrence matrix $M$, where rows represent repositories (all repositories from all versions) and columns represent libraries. A matrix element $M_{R,L}$ equals to one if repository $R$ directly depends on library $L$, and zero otherwise. 
Then, we apply SVD to decompose $M$ into a product of three matrices $U$, $\Sigma$, and $V$. $\Sigma$ is a matrix with singular values of $M$ on its main diagonal, rows of $U$ are embeddings of repositories, and columns of $V$ are embeddings of libraries. Initially, the dimensionality of embeddings equals to the total number of repositories and libraries in the dataset, respectively. In order to get small and meaningful vectors, we drop the components that correspond to the smallest singular values, which leaves us with embeddings of fixed size $d$ for both libraries and repositories. In this study, we used the implementation of SVD from scikit-learn~\cite{scikit-learn}. We evaluated different dimensionalities for this task and decided on using 32-dimensional embeddings, more details on their comparison are presented in \Cref{sec:compare}.

\secpart{Exploring embedding semantics}
\label{sec:clusters}

In order to evaluate whether the built embeddings of libraries contain semantic information, we clustered them with the K-means algorithm~\cite{kmeans} and explored the meaningfulness of the constructed clusters. K-means works as follows. Firstly, we fix the number of clusters $K$ and the algorithm chooses $K$ random cluster centers. In the main stage of K-means, we iteratively assign items to the closest center to form clusters, and recompute cluster centers as the mean of items in each cluster. The implementation of K-means was also taken from scikit-learn~\cite{scikit-learn}.

\begin{figure}[h]
  \centering
  \includegraphics[width=\columnwidth]{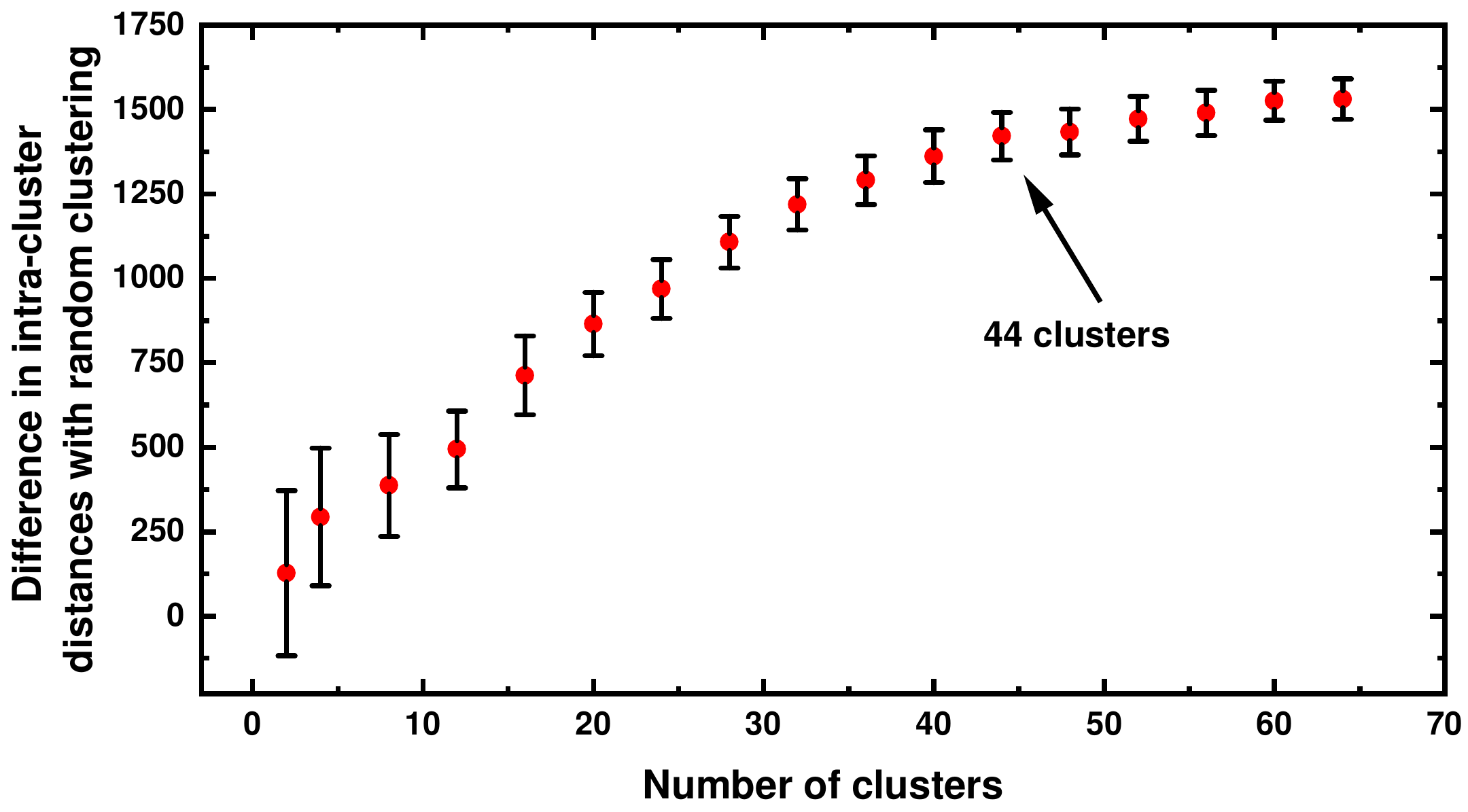}
  \caption{The dependence of the difference of intra-cluster distances between the built cluster partitions and random clusterings of the same size on the number of clusters when clustering the library embeddings with the K-Means algorithm. The arrow indicates the saturation point.}
  \label{fig:gap}
\end{figure}

To choose $K$, we used the \emph{gap statistic} technique~\cite{gapStatistic}. We varied the number of clusters from 2 to 64, and for each cluster partition computed the intra-cluster distance (\ie the total distance from cluster vectors to the cluster center). Then we computed the difference of intra-cluster distances between the built cluster partitions and random clusterings of the same size. With the increase of the number of clusters, the difference increases, because more clusters allow to divide the data more granularly. \Cref{fig:gap} shows the resulting dependence. We conclude that the difference begins to stabilize at $K$ equals 44, hence we used this value to further study the vector space.

\begin{figure*}
  \centering
  \includegraphics[width=\textwidth]{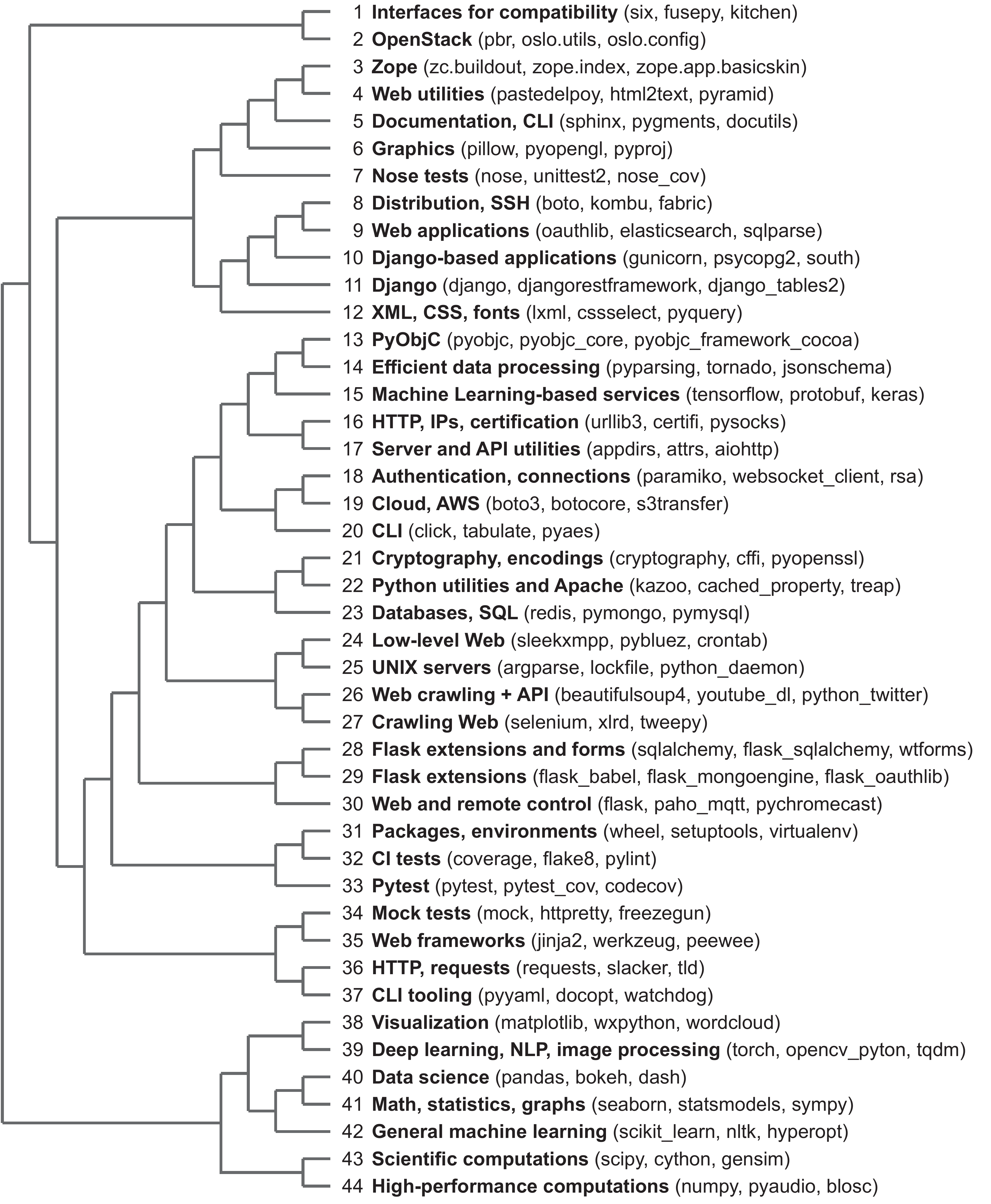}
  \caption{The dendrogram of the constructed clusters.}
  \label{fig:dendrogram}
\end{figure*}

To analyze the clusters, for each cluster, we manually inspected the most frequent libraries assigned to it, as well as repositories, for which most of the dependencies belong to this cluster. For convenience, we gave a short name to every cluster. We also constructed a dendrogram shown in \Cref{fig:dendrogram} that illustrates the process of merging similar clusters into larger groups. To build the dendrogram, we used the scikit-learn~\cite{scikit-learn} implementation of agglomerative clustering scheme with cosine distance and average linkage.

It can be noticed that both the clusters and their relations to each other are semantically sound. In the vast majority of cases we had no trouble understanding and naming them. As the dendrogram suggests, the clusters are divided into three categories at the root.

The first category consists of clusters 1 and 2, which are the furthest away from the others. Cluster 1 is very broad and domain-independent, it contains libraries that relate to compatibility, for example, of Python versions. Cluster 2, in contrast, is very specific and contains libraries used by the \textit{OpenStack} organization on GitHub.

The second large group of clusters contains Clusters 38--44 and can be named \textit{Science and Machine Learning}. The presence of such a group can be expected, since Python is extremely popular in scientific applications. In this group we can see two clusters that deal with calculations, clusters that deal with data analysis, statistics, and machine learning, as well as libraries for plotting and visualizing data. It is within this group of clusters that we can see the most popular data science libraries like \textit{numpy}, \textit{scipy}, and \textit{pandas}, the most popular libraries for machine learning and deep learning like \textit{torch} and \textit{scikit\_learn}, and the most popular tools for plotting graphs: \textit{matplotlib} and \textit{seaborn}.

Finally, the largest group of clusters contains Clusters 3--37 and might be called \textit{Web and applications}. This group itself divides into several other noteable sub-groups.  

Aside from analyzing high-level information about the state of the Python open-source ecosystem (\eg the presence of different domains and their relative sizes), the constructed dendrogram can also provide more detailed insights. As an example of the latter, \textit{Tensorflow} --- one of the most popular machine learning frameworks --- belongs to Cluster 15, while most of the other ML libraries belong to Clusters 37--44. After a more detailed analysis of the repositories that depend on each group of clusters, we found that \textit{Tensorflow} is a more frequent choice for online services that provide ML-based functionality (\eg face recognition or sentiment analysis in the cloud). This insight also explains the relatively small distance between the Tensorflow's cluster and libraries related to AWS.

We conclude that the constructed library embeddings are indeed meaningful and carry enough semantic information for both high-level descriptive analysis and providing interesting insights about the libraries. It leaves us with a strong basis to further explore their practical applicability.
\section{Recommending relevant libraries}
\label{sec:recom}

In order to prove the usefulness of the constructed library embeddings in the SE domain, we employ the task of dependency recommendation that was previously used by Theeten et al.~\cite{import2vec}. The rationale behind it is a rapid growth in the number of open-sourced Python libraries: as the number of available frameworks and tools grows, it is important to keep up with the innovations and explore relevant libraries used in similar projects. 

Unfortunately, the previous work did not introduce a proper way to compare the recommender systems. Thus, for the evaluation of the proposed approach and the comparison of different models, we need a benchmark with a fixed definition of a correct and an incorrect suggestion. In our work we compiled such a benchmark using the historical data of the previous versions of requirements files in our dataset.

However, such a benchmark is still synthetic to some degree and does not fully capture the relevance and the usefulness of the recommendation. To combat this, we also conduced a user study about the suggested libraries. In this section, we will describe the recommender systems that we used and both of the mentioned evaluations.

\subsection{Recommender systems}

\begin{figure*}
  \centering
  \includegraphics[width=\textwidth]{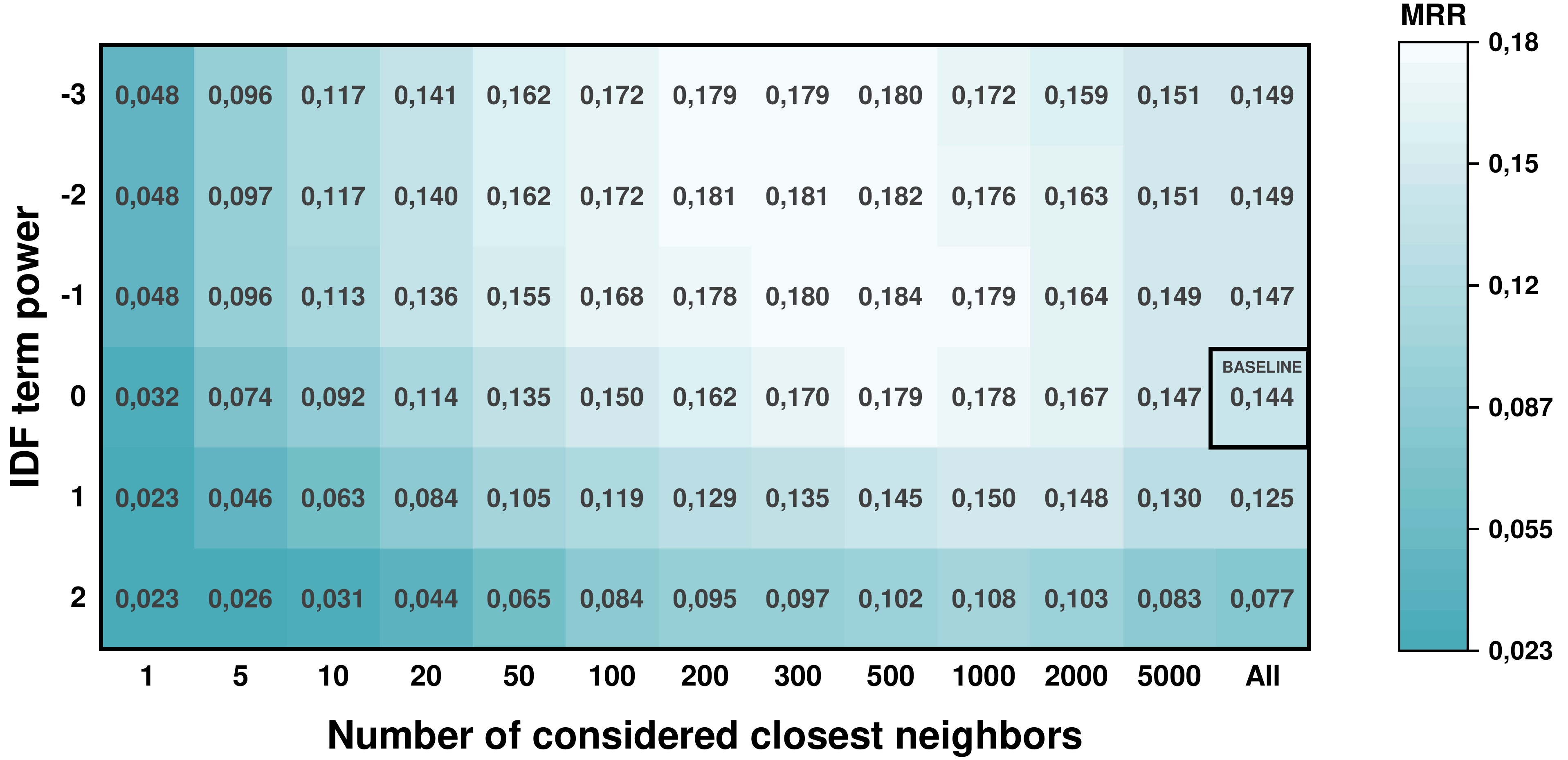}
  \caption{The dependence of the \libraries model's MRR on the model parameters with the cosine similarity power fixed at 0. The value in the square corresponds to the \baseline.}
  \label{fig:idf}
\end{figure*}

Firstly, we suggest a recommender system based on the constructed library embeddings. The general idea of the the approach is to find repositories similar to the given one, look at their dependencies and suggest those that are not already in the query project. To find the closest repositories, we need to assign vectors to them as well.

We assign vectors to repositories as a mean of embeddings of their dependencies. Then, given a repository $R$, we find $K$ repositories closest to $R$ according to cosine distance, where $K$ is a parameter chosen empirically. For each library $L$ present in these repositories that is not present in $R$, we compute their relevance score as following:

\begin{equation}\label{eq:1}
    score_{L} = idf(L)^\alpha \sum\limits_{r \in \text{Nearest}(R, L)} (1-dist(r, R))^\beta,
\end{equation}

where $\alpha$ and $\beta$ are hyperparameters, $dist$ is cosine distance, $idf$ is inverse document frequency of the library in the entire dataset~\cite{Jones72idf}, and summation is over repositories closest to $R$ that also contain $L$. The intuition behind this formula is that we assign higher score to libraries that frequently occur among repositories similar to $R$, and reweigh the score given the library's popularity represented by $idf$. Finally, the system recommends libraries with the highest relevance score.

From here on out, we will refer to the system that represents repositories as a mean of their libraries as \textit{Repositories by Libraries Embeddings} (\libraries). To verify the adequacy of this system, we compare it to three alternative approaches:

\begin{itemize}
    \item \textit{Direct Repository Embeddings} (\direct). Instead of assigning the mean of dependency embeddings to each repository, we could use the embedding directly computed in SVD (see \Cref{sec:svd}). A downside of this approach is its inability to handle new repositories: in order to build a vector for them we would need to retrain the entire SVD once again. Equation~\ref{eq:1} stays the same.
    \item \jaccard. We could directly compute similarity of repositories as a Jaccard distance~\cite{jaccard} between sets of their dependencies. While it might be more intuitive, searching for similar repositories according to Jaccard distance requires the computation of a set intersection with each repository in the dataset, while for cosine distance we can make the search asymptotically faster using K-D Trees~\cite{kdtree}. Equation~\ref{eq:1} stays the same.
    \item \baseline. As a baseline, we use the recommendation of the most popular libraries among all the repositories. This recommendation is the same for any repository and pays no attention to the query project.
\end{itemize}

\subsection{Creating the benchmark}

To create a benchmark for comparing the models and estimate their best parameters, we decided to utilize the obtained historical data, namely, the requirements files from different years (see \Cref{table:dataset}). The idea is to compile the projects that have actually added some libraries to them and try to suggest these libraries. 

More specifically, among the projects from 2011--2019 (\textit{i.e.}, all projects that have the \textit{next} version), we selected all projects that have:

\begin{itemize}
    \item added at least one new library in the next version;
    \item did not change (add or remove) more than 10 libraries.
\end{itemize}

The second limitation removes only a handful of candidates, but by manual checking we found a few cases where between versions the project changed \textit{completely} (moved to entirely different domain), and trying to predict this is unreasonable.

Overall, the benchmark consisted of 4,678 projects. For example, if project \textbf{X} had dependencies [\textbf{A}, \textbf{B}, \textbf{C}] in 2016 and dependencies [\textbf{A}, \textbf{B}, \textbf{D}] in 2017, the task would be: given the repository with dependencies [\textbf{A}, \textbf{B}, \textbf{C}] and considering only projects from 2016, suggest the dependency \textbf{D}.

When evaluating models on the benchmark, we only considered the neighbors from the projects in the same temporal slice as the target project. The same goes for the baseline --- the list of the most popular libraries is calculated for the year of the query project. All of this is done to prevent the possible cases of looking into the future. 

For each of the models with each parameter configuration, we calculated the metrics commonly used for the recommendation systems: precision among first K suggestions, recall among first K suggestions, and MRR (Mean Reciprocal Rank).

\subsection{Comparing the models on the benchmark}
\label{sec:compare}

All the tested models, apart from the \baseline, allow us to vary three parameters from Equation~\ref{eq:1}: $\alpha$ --- the power of the IDF term, $\beta$ --- the power of the cosine similarity term, and $K$ --- the number of the closest repositories to consider.

\begin{itemize}
    \item We varied the power of the IDF term from -3 to 2. When this term is raised to the negative power, the model facilitates recommending popular libraries, and when this term is raised to the positive power, the model facilitates suggesting more obscure, rare libraries. If the power is equal to 0, the popularity of the library is not considered. 
    \item We varied the power of the cosine similarity term from 0 (ignoring the distance altogether) to 2.
    \item We varied the amount of the nearest neighbors from 1 to all of the repositories.
\end{itemize}

It should also be pointed out that using any of the models with the terms' powers equal to 0 and considering all the repositories as neighbors results in the \baseline.

We discovered that the power of the similarity term does not significantly impact the metrics. However, the IDF term power and the number of considered neighbors both did. \Cref{fig:idf} shows the heatmap of the MRR value with a fixed similarity term power of 0. It can be seen that negative values of the IDF term power demonstrate better results. A negative IDF term power facilitates the model to predict more popular libraries, and since the benchmark was built on historical data, such libraries were naturally added more often. It can also be seen that the best results are obtained when considering a moderate number of neighbors (200--500), which backs up the idea of using similar projects to make the suggestions. Finally, the heatmap shows that the \libraries model with optimal parameters significantly outperforms the \baseline.

MRR of the \direct model behaves very similarly, with the best results obtained at the same parameters as the \libraries model. Probably, this has to do with the similar ways in which the embeddings were obtained.

As for the \jaccard model, its relation to the powers of the terms is the same, the main difference is that the best results are obtained when considering 100 closest neighbors. While \libraries and \direct models utilize semantic similarity of dependencies via embeddings and can identify similar projects even when they have no dependencies in common, \jaccard model relies on the explicit intersection of dependency sets. Thus, the number of meaningful neighbors to consider is smaller in the \jaccard case.

\begin{table}[h]
\centering
  \caption{The best results of different models on the benchmark. $\alpha$ is the power of the IDF term, $\beta$ is the power of the cosine similiarity term, and $K$ is the number of the considered closest neighbors.}
  \label{table:benchmark}
  \begin{tabular}{l c c c c c c}
    \toprule
    \multirow{2}{*}{\textbf{Model}} & \multicolumn{3}{c}{\textbf{Parameters}} & \multirow{2}{*}{\textbf{Prec@1}} & \multirow{2}{*}{\textbf{Rec@10}} & \multirow{2}{*}{\textbf{MRR}}\\
    \cmidrule(lr){2-4}
                                    & $\alpha$ & $\beta$ & $K$                      &                               &                                &\\
    \midrule 
    \libraries & -1 & 2 & 500 & 0.104 & 0.184 & 0.185 \\
    \direct & -1 & 2 & 500 & \textbf{0.105} & 0.193 & \textbf{0.189} \\
    \jaccard & -1 & 2 & 100 & 0.104 & \textbf{0.195} & 0.186 \\
    \baseline & --- & --- & --- & 0.075 & 0.140 & 0.144 \\
    \bottomrule
  \end{tabular}
\end{table}

\Cref{table:benchmark} shows the overall results of different models with the best parameters. All numbers are relatively low, because the task is inherently very difficult, but it can be seen that all three models perform similarly and significantly better than the \baseline. This means that considering the closest repositories is useful for the task, and each of the proposed models is potentially usable. In practice, the \libraries model has significant advantages over the others. As previously mentioned, the \jaccard model requires significantly more calculations as the number of repositories grows, and the \direct model is not easily expandable to new projects. 

We built a simple CLI tool that runs the \libraries model on a user-defined input. It requires only the \textit{requirements.txt} file of the target project and uses pre-calculated embeddings for rapid searching of the nearest neighbors. 

Having a benchmark also allowed us to compare different dimensionalities of the SVD embeddings. We experimented with ten different dimensionalities of the form $2^n$ with $n$ ranging from 0 to 9. We considered both \libraries and \direct embeddings, and noticed that the quality of predictions saturates at the dimensionality of 32. Increasing the number of dimensions further does not improve the results on the benchmark. Considering also the semantic meaningfulness of the obtained embeddings (See \Cref{sec:clusters}), we believe that 32 dimensions is optimal for SVD in this task.

\subsection{User study}

The historical benchmark is a useful tool for comparing the models and tuning hyperparameters. However, the ranging metrics that we discussed do not guarantee practical value for developers because of the synthetic nature of the compiled benchmark. The goal of the recommendation system is to provide relevant and useful suggestions, and since these concepts are inherently subjective, the evaluation should be carried out on people. For this reason, we also conducted a user study.

Firstly, we needed to select repositories for the evaluation. As discussed in \Cref{sec:clusters}, libraries can be assigned to broader domains by their topic. To compile the user study, we used five domains of Python projects mentioned by Lin et al.~\cite{lin2016empirical}: NLP, Web, Media, Data processing, and Scientific Calculations. As target repositories, we chose one repository per domain from our dataset, specifically, from the 2020 temporal slice that was not present in the benchmark and, in a sense, has no \textit{ground truth}. Let us now describe the repositories in each domain.
\begin{table*}[]
\centering
\caption{The results of the user study. The answers to all three questions for each of the library predicted by a given model were averaged for all the participants. The last three columns show the average of the results for all five repositories.}
\label{table:user_average}
\begin{tabular}{lccc c ccc c ccc c ccc c ccc c ccc}
\toprule
\multicolumn{1}{l}{\textbf{Domain}} & \multicolumn{3}{c}{\textbf{NLP}} && \multicolumn{3}{c}{\textbf{Web}} && \multicolumn{3}{c}{\textbf{Media}} && \multicolumn{3}{c}{\textbf{DP}} && \multicolumn{3}{c}{\textbf{SC}} && \multicolumn{3}{c}{\textbf{Average}} \\ \cmidrule(lr){2-4} \cmidrule(lr){6-8} \cmidrule(lr){10-12} \cmidrule(lr){14-16} \cmidrule(lr){18-20} \cmidrule(lr){22-24}
\textbf{Question}                   & \textbf{I}      & \textbf{II}     & \textbf{III}   && \textbf{I}      & \textbf{II}     & \textbf{III}   && \textbf{I}       & \textbf{II}     & \textbf{III}    && \textbf{I}      & \textbf{II}    & \textbf{III}   && \textbf{I}      & \textbf{II}    & \textbf{III}   && \textbf{I}        & \textbf{II}      & \textbf{III}     \\ \midrule
\textbf{By libraries}               & 3.4   & 2.7   & 3.3  && 3.3   & 2.3   & 3.1  && 1.7    & 1.5   & 1.7   && 3.4   & 2.8  & 3.2  && 2.7   & 2.1  & 2.5  && 2.9     & 2.3    & 2.8    \\
\textbf{Direct}                     & 3.2   & 2.5   & 3.2  && 3.1   & 2.2   & 3.0  && 1.8    & 1.7   & 2.0   && 3.6   & 2.9  & 3.6  && 2.2   & 1.8  & 2.1  && 2.8    & 2.2    & 2.8    \\
\textbf{Jaccard}                    & 2.5   & 2.1   & 2.6  && 1.8   & 1.6   & 1.8  && 1.5    & 1.5   & 1.6   && 4.0   & 3.3  & 3.9  && 2.0   & 1.6  & 1.9  && 2.4     & 2.0    & 2.3    \\ \bottomrule
\end{tabular}
\end{table*}
\begin{itemize}
    
    \item \textbf{NLP}: Seq2seq Chatbot~\cite{seq2ser_chatbot} -- a lightweight implementation of a chatbot.
    
    \item \textbf{Web}: Octopus~\cite{octopus} -- a server for accessing PowerShell via HTTP(S).
    
    \item \textbf{Media}: Thumbnail Generator~\cite{thumbnail_generator} -- a tool for generating thumbnail sprites from a video.
    
    \item \textbf{Data Processing (DP)}: Fuel~\cite{fuel} -- a framework for processing datasets: working with popular formats, iterating, editing.
    
    \item \textbf{Scientific Calculations (SC)}: SciKit Kinematics~\cite{scikit_kinematics} -- a collection of functions for working with 3D kinematics, like rotation matrices and quaternions.

\end{itemize}

In our user study, we evaluated three best performing models from \Cref{table:benchmark}. We did not evaluate the baseline (the predictions of which are analogous to libraries in \Cref{table:top10}) in order not to diffuse the user study with the same general suggestions, but rather focus on different domain-specific suggestions. 

For each of the five projects, we used the best performing models to generate Top-5 most relevant suggestions and merged all of them into a single list in case of repetitions. All participants were shown all the suggestions for a given project in random order and were asked three questions about each suggestion, with the answers being a scale from 1 to 5:

\begin{enumerate}[I]
    \item \textbf{Is this suggestion relevant to the topic of the project?} 
    
    \textit{1 -- not relevant at all, 5 -- very relevant.}
    \item \textbf{Would you use this library in this project?} 
    
    \textit{1 -- no, the library either does not bring any value or requires too much technical work to use in this project, 5 -- yes, it is valuable and does not require significant efforts to start using in the present project.}
    \item \textbf{Would you consider using it if you started another similar project?} 
    
   \textit{ 1 -- no, the library cannot bring any value to similar projects, 5 -- yes, the library can be useful as an addition or as a replacement of one of the currently used tools.}
\end{enumerate}

In total, \numSurvey experienced Python developers participated in our study. The averages of their answers are presented in \Cref{table:user_average}. Several observations can be made from them.

First of all, it can be seen that the results are very unequal between domains. For three domains (NLP, Web, and Data Processing), embeddings-based models demonstrated the average scores in questions I and III better than 3, indicating the overall relevant and potentially useful recommendation. At the same time, the results for the other two domains (Media and Scientific Calculations) are noticeably and significantly worse. 

Even though in our work we only have one repository per domain, during our preliminary experiments we have seen similar cases for other repositories in these domains. \Cref{fig:dendrogram} sheds the light on  this discrepancy. All the domains that showed better results are well presented in the clusters, and at the same time, Media and Scientific Calculations are both mentioned in just one cluster, which makes them under-represented. The models consider 100 or even 500 closest repositories, so if there are significantly less related projects, the unrelated projects might bring up unrelated suggestions. This is exactly what happened to Thumbnail Generator: since there are few Media-related projects in our dataset, the closest projects contained a lot of weakly related repositories, and the suggestions of all three models contained just generally popular libraries, some of which you can see in \Cref{table:top10}: \textit{numpy}, \textit{six}, \textit{requests}, etc. 

Even though the two under-represented domains bring down the overall average, the approach shows good results for the domains that are popular in Python. While \Cref{table:user_average} shows the \textit{average} results of all the suggested libraries, \Cref{table:user_best} shows the score of the \textit{best} library for every model in regard to Question III about usefulness in a similar project. It can be seen that for the three well-represented domains, Top-5 suggestions of our embeddings-based models contained libraries that were voted overwhelmingly useful. 

\begin{table}[h]
\centering
\caption{The highest score of a single best library predicted by each model in regard to question III.}
\label{table:user_best}
\begin{tabular}{@{}lccccc@{}}
\toprule
\textbf{Domain}       & \textbf{NLP}           & \textbf{Web}           & \textbf{Media}         & \textbf{DP}            & \textbf{SC}            \\ \midrule
\libraries & \textbf{4.3} & \textbf{4.0} & \textbf{2.7} & \textbf{4.6} & \textbf{3.0} \\
\direct       & \textbf{4.3} & \textbf{4.0} & \textbf{2.7} & 4.3          & 2.9          \\
\jaccard      & 3.6          & 3.2          & 2.3          & \textbf{4.6} & 2.9          \\ \bottomrule
\end{tabular}
\end{table}

Another conclusion that can be drawn from \Cref{table:user_average} and \Cref{table:user_best} is that on average, embeddings-based models demonstrate better results than the \jaccard model. The \libraries model was able to suggest all the libraries that received the highest scores in all questions. Overall, this can be viewed as another advantage of using embeddings for a given task.

Finally, \Cref{table:user_average} demonstrates another important implication: for all the models in all the domains, and on average, the score in Question II is lower than the scores in Questions I and III. This means that using the suggestions in the \textit{given} project is more difficult than in a \textit{similar} project, and high relevance does not guarantee the ability to use the library right away. 

To investigate this further, we asked the participants to comment on why they found certain libraries relevant but not useful. There were two most common answers. Firstly, all the suggested libraries are very popular and therefore the decision to not include them might be voluntary. This is not unusual, because, as we discussed before, all the best-performing models use the negative power of the IDF term, and therefore suggest more popular libraries. Secondly, the suggested libraries can be \textit{analogous} to the ones that are already in the project: even if they might bring additional value, it might be difficult to integrate them to a mature project.

Overall, the demonstrated results show the potential of the proposed approach. It is important to keep in mind how little information this method uses to make the decisions. However, the results of the user study point to a fundamental discrepancy between the relevance and the usefulness of the suggestions. 

\subsection{Exploration of libraries}

As a final part of our study, we decided to investigate this problem a bit deeper. Let us demonstrate the discrepancy between the relevance and the usefulness of the suggestions by considering the first of our studied projects, the ChatBot. Its dependencies are somewhat typical for a Machine Learning project and include \textit{numpy} --- a library for scientific calculations, \textit{scikit\_learn} --- a library for data analysis, and \textit{tensorflow} --- one of the most popular machine learning frameworks.

Now let us look at the recommendations made by the \libraries model. The top suggestions are relevant to the topic of the project and include \textit{torch} -- another popular machine learning framework, \textit{pandas} -- another data analysis tool, and \textit{matplotlib} -- one of the most popular libraries for plotting and visualization. Naturally, such suggestions are more relevant than, for example, the \baseline, and can be helpful in search engines or package management services, where they can guess what the developer might search for. However, if we are looking for libraries that can \textit{add something new} to the project, these suggestions turn out to be less useful: it is doubtful that developers will rewrite the application in another framework.

If we go back and look at the distributions of libraries demonstrated in \Cref{sec:distribution} and in \Cref{fig:dependencies}, they are very unequal, with a handful of very popular libraries and a long ``tail'' of rare ones. It might be of interest to suggest such rare libraries to allow the developers actually discover some potentially \textit{useful} libraries. We refer to this mode of recommendation as \emph{exploration}.

The parameters of our models can facilitate such exploration. Firstly, and more obviously, we can increase the power of the IDF term, therefore incentivizing the models to predict rarer libraries. Secondly, we can decrease the number of considered closest neighbors, therefore making the possible suggestions more precise.

Choosing the parameters of models for such a task is very difficult and requires complex manual evaluation. We leave this for the future work, but demonstrate a couple of hand-picked examples for the \libraries model with the the IDF term power of 3 and the number of closest neighbors of 50.

If we use these settings for the Chatbot, the first five suggestions include \textit{keras-bert} -- an implementation of the BERT Language model, and two libraries that were encountered only once in the entirety of our dataset: \textit{lemminflect} -- a library for lemmatizing English words and \textit{language-tool-python} -- a grammar and spelling checker. These libraries relate closely to the topic of the repository and there is a very good chance that the developer has not heard of them. 

Of course, boosting up very rare libraries is more risky, and other suggestions can be totally irrelevant. There are possible ways of mitigating this, for example, clustering the repositories and considering not just the closest projects, but only the projects within the same cluster. Overall though, this approach may be useful for discovering interesting libraries that are otherwise shadowed by the most popular ones.
\section{Threats to Validity}
\label{sec:threats}

The general large-scale nature of our study gives way to certain treats to validity.

\textbf{Parsing}. When parsing the requirements, we took into account the possible upper and lower case, as well as the fact that the symbols ``-'' and ``\_'' are interchangeable in the \textit{pip} formatting. However, there exist specific cases where the same library can have different aliases. For example, a popular Web scraping library \textit{beautifulsoup4} is often called by its short name \textit{bs4}, and the developers of the library have uploaded a dummy placeholder package to the abbreviated name that ensures the downloading of the same library and prevents possible misunderstandings. If there are other such cases in the dataset, our approach will not be able to match such aliases, however, such cases are rare.

\textbf{Python versions}. In this work, we also do not differentiate between Python3 and Python2 and can therefore suggest deprecated packages. However, all the most popular libraries are well-maintained and do not have this problem, whereas the exploration suggesting mode is intrinsically unstable. Moreover, when suggesting relevant libraries, only the project versions from the same year are taken into account, which can also prevent recommending old libraries.

\textbf{Indirect dependencies}. Some libraries have dependencies of their own that do not need to be in the requirements files or be explicitly used but are installed automatically. For this reason, there is a possibility that our approaches might suggest libraries that are already installed and used under the hood in the developer's project. However, the goal of suggesting libraries is to advise the developer to take a look at the library, and even if it is used under the hood, the developer might not know it closely, and the suggestion might still be useful.

\textbf{Choice of projects}. In our user study, we only selected one project per domain. This has to do with the fact that labeling several predictions for several models for each repository constitutes a lot of work. However, this only directly impacts one of our observations that relates to the differences between domains, while the other ones deal with the averaged data of all five projects. We leave further extended experiments for future work.
\section{Conclusion \& Future Work}
\label{sec:conclusion}

In this paper, we propose an idea of creating lightweight embeddings of libraries and projects based on their co-occurrence. We gathered a dataset consisting of requirements files of 7,132 Python projects and pulled up their versions from the years 2011--2020. Based on this data, we obtained dense 32-dimensional vectors of libraries by applying SVD to the co-occurrence matrix of projects and libraries.

We evaluated the semantic meaningfulness of the extracted embeddings by clustering them with K-means and manually inspecting the clusters. The clusters themselves and their relative positioning demonstrate that the embeddings carry with themselves a lot of meaning and can be used for making relevant suggestions.

Based on this, we developed an approach to suggest relevant libraries to the developers of a given project based on the project's dependencies. The idea of the approach is to search for similar repositories and suggest their dependencies for consideration. To identify similar repositories, we tried embedding-based approaches and Jaccard similarity between the dependency sets. We also used a baseline of simply suggesting the most popular libraries overall.

To test and compare models, we compiled a benchmark based on the historical data from our dataset. The models suggested libraries to the projects in a given year, and tried to guess the actually added dependencies. The baseline demonstrated the MRR of 0.144, while all similarity-based models demonstrated similar results, with the highest MRR of 0.189. We also implemented the approach based on library embeddings as a CLI tool for others to try.

To complement the comparison of models on a synthetic benchmark, we conducted a user study. We used the similarity-based models to suggest libraries for five different projects, and asked the participants to evaluate their relevance and usefulness. The study showed that the suggestion quality varies between the project domains. Also, it turns out that the recommendation of popular libraries might not actually bring any new knowledge to the developer. Therefore, we introduced another mode for our recommender system called \textit{exploration} that facilitates the system to score rare libraries higher. 

Overall, we believe that the proposed embeddings might be of interest and of use for various tasks. They are very simple to obtain, require no text or code analysis, but at the same time allow us to differentiate between different domains. There is a number of ways to continue this research:

\begin{itemize}
    \item It is possible to test the proposed approach for different languages and package managers.
    \item There is a need to study the process of obtaining embeddings in greater detail, and to compare different approaches to this task.
    \item It is of interest to carry out a user study of the recommendation models on a larger scale, considering different ways of suggesting relevant libraries, including the exploration mode.
\end{itemize}

We believe that it is very important to consider various sources of information when analyzing the software ecosystem, and hope that our work can be helpful in this regard.

\section*{Acknowledgments}

We would like to thank Viktor Tiulpin for his help and expertise with GHTorrent and data collection.

\bibliographystyle{ieeetran}
\balance
\bibliography{cites}

\end{document}